\input{aipcheck}

\documentclass[
    ,final            % use final for the camera ready runs
  ]
  {aipproc}

\def\lp {\left( }
\def\rp {\right) }
\def\lb {\left[ }
\def\rb {\right] }
\def\lc {\left\{ }
\def\rc {\right\} }
\def\ra {\rangle }
\def\la {\langle }
\def\ni {\noindent}
\def\nn {\nonumber}

\def\rar {\rightarrow}

\def\beq{\begin{equation}}
\def\eeq{\end{equation}}
\def\bea{\begin{eqnarray}}
\def\eea{\end{eqnarray}}

\def\dr {\partial }

\def\cO{{\cal{O}}}

\def\d{\delta}
\def\D {\Delta}
\def\e{\epsilon}

\def\m{\mu}

\def\p {\pi}

\def\s{\sigma}

\def\sp {\!+\!}
\def\sm {\!-\!}

\def\cd {\!\cdot\!}

\def\ba {\mbox{\boldmath $a$}}

\def\bFF {\mbox{\boldmath $F$}}

\def\bk {\mbox{\boldmath $k$}}

\def\br {\mbox{\boldmath $r$}}

\def\bsig {\mbox{\boldmath $\sigma$}}

\def\bnb {\mbox{\boldmath $\nabla$}}

\def\bq {\mbox{\boldmath $q$}}

\def\bv {\mbox{\boldmath $v$}}
\def\bx {\mbox{\boldmath $x$}}

\layoutstyle{6x9}

\begin{document}

\title{Three-nucleon interactions: dynamics}


\author{M. R. Robilotta}
{  address=
{Instituto de F\'{i}sica, Universidade de S\~ao Paulo,
S\~ao Paulo, SP, Brazil} }

\begin{abstract}
A discussion is presented of the dynamics underlying three-body nuclear forces, 
with emphasis on changes which occurred over several decades.
\end{abstract}

\maketitle

%%%%%%%%%%%%%%%%%%%%%%%%%%%%%%%%%%%%%%%%%%%%
%% MAINMATTER
%%%%%%%%%%%%%%%%%%%%%%%%%%%%%%%%%%%%%%%%%%%%

%--- 1 ---------------------------------------------------------------------
\section{a warning as introduction}

In this work, I sketch the development of nuclear three-body
forces over a rather large span of time and concentrate
narrowly on dynamics.
As a consequence, many important subjects are completely omitted, 
especially those concerning experimental facts and facilities, 
phenomena and calculation techniques. 
Even within the restricted domain of dynamics, many relevant contributions
could not be mentioned or discussed, owing to limitations of space.
A more comprehensive list of references can be found 
in a previous related presentation\cite{R87}.

This work is definitely {\bf not} a historical reconstruction,
even if many developments are presented in a rough chronological order.
The various sections are organized around themes possessing 
a kind of internal coherence, which yield something like
a research mood. 
I have tried to convey these moods and to suggest that they change 
from one period to another.
With that purpose in mind, most technicalities were avoided 
and many original quotations were made, 
indicated by both quotation marks and italic fonts. 

%--- 2 -----------------------------------------------------------------------
\section{Classical three-body forces}

Classical Electrodynamics, as represented by Maxwell's equations,
is {\bf the} paradigm for describing particle interactions mediated by fields.
In Electrostatics, the substructure dealing with systems of charges at rest,
the Coulomb law holds and the potential energy $U$ is a meaningful concept.
In the case of just a pair of particles $1$ and $2$, the potential energy has 
the form $U_{12}= K Q_1 Q_2/r_{12}$. 
When several particles are present, one usually invokes the linear 
superposition principle in order to generalize this result.
Accordingly, the potential energy of a system with three particles would 
be given by
\beq
U=U_{12}+U_{23}+U_{31} \;.
\label{e.1}
\eeq

The manifestation of the superpositon principle in this result is the fact 
that the potential energy of the system is {\bf not} written as  
\beq
U'=U'_{12}+U'_{23}+U'_{31} \;,
\label{e.2}
\eeq

\ni
with $U'_{ij}\neq U_{ij}$.
The principle implements the idea that the interaction between 
particles $1$ and $2$ is completely independent 
of the presence of the third particle.
This idea is based on a tacit assumption, namely that the charge 
distributions of the particles involved are rigid and do not change
in the interaction process.
In Nature, however, this kind of condition is satisfied only in a few 
special cases.

Quite generally, if a particle has parts, it must also contain internal 
forces, responsible for the binding of these parts.
As a necessary consequence, when the particle is placed in the presence 
of intense enough external forces, deformations can be produced.
Only elementary particles, which by definition do not have parts, 
cannot be deformed by external interactions.
They also cannot have size, for this would imply the existence of parts.
Elementary particles must therefore be point-like, 
with the word point used in its mathematical sense. 

Going back to Electrostatics, one sees that the linear superposition 
principle can hold only for truly elementary particles.
In the case of systems, such as molecules, atoms or nucleons,
one indeed has $U'_{ij}\neq U_{ij}$ and eq.(\ref{e.1}) corresponds 
to an approximation.
The convenience of this approximation depends on the problem considered and, 
in many instances, precision demands the inclusion of corrections.
These are usually taken into account by means a {\bf three-body potential} 
$W$, such that $W\equiv U'-U$.
This definition ensures that the full interaction is suitably recovered by the 
joint use of $U$ and $W$.
Physically, this class of three-body forces is associated with deformations
of the interacting objects.  

Three-body interactions can also be generated by another kind of mechanism,
which depends on the complete Maxwell's picture.
When two charged particles $1$ and $2$ are free to move, both electric and 
magnetic interactions occur.
The force acting on one of the particles depends then on the relative 
distance $\br_{12}$ and on the individual velocities $\bv_1$ and $\bv_2$.
Moreover, as the particles are accelerated, interaction through 
radiation is also possible.
Even when the velocities involved are not very high, the complete force
$\bFF_1$ acting on particle $1$ is a complicated function of the form
\beq
\bFF_1= \bFF_1(\br_{12}, \,\bv_1, \,\bv_2, \,\ba_2)\;.
\label{e.3}
\eeq 

If a third charge is brought into the system, it will give rise to new
forces over particles $1$ and $2$.
In particular, the new force over charge $2$ will induce a change 
$\ba_2 \rightarrow \ba'_2$ in its acceleration.
The inclusion of this effect into eq. (\ref{e.3}) promotes an 
{\bf indirect} influence of the new particle over charge $1$,
which also corresponds to a three-body force.

%--- 3 ----------------------------------------------------------------------
\section{Early three-body forces}

The first work to deal with quantum three-body forces was entitled
{\em "Many-Body Interactions in Atomic and Nuclear Sytems"} and
produced in 1939, by Primakoff and Holstein\cite{PH39}.
They begin by considering the second classical mechanism mentioned 
in the previous section, associated with eq.(\ref{e.3}),
and then move on to quantum electromagnetic interactions.
The conclusion is reached that 
{\em "classical many-body potentials"} 
correspond to the case in which an electron 
{\em "simultaneously emits two virtual quanta"}.
On the other hand, they note that the process in which
{\em "one electron emits two virtual light quanta in succession"},
gives rise to 
{\em "specifically quantum-mechanical many-body potentials".}
The same kind of interactions are considered in the section 
{\em "Mesotron Field Theory of Nuclear Interactions"},
with the light quanta replaced by vector Yukawa particles.
By estimating the typical sizes and velocities of atomic and nuclear systems, 
they infer that 
{\em "the description of the electromagnetic interaction of electrons
in atomic systems, by means of action-at-a-distance two-body potentials
is an extraordinarily good approximation [...]"},
whereas 
{\em "The usual description of nuclei in terms of two-body potentials 
cannot [...] be considered satisfactory, except in the case of the 
deuteron"}. 

An extension of this discussion, together with an application to the 
trinucleon system, were the object of another paper, published 
in the same year\cite{J39}.

%--- 4 -----------------------------------------------------------------------
\section{Early strong theory}

The discoveries\cite{Pi} of the charged pions, in 1947, and of their neutral 
counterpart, in 1948, were followed by a consistent effort aimed at 
producing a theory of nuclear forces. 
This research program relied on the idea of a range expansion,
outlined by Taketani, Nakamura and Sasaki\cite{TNS51}, in 1951.
Soon afterwards, field theory was being employed in nuclear calculations,
with great vigour.
 
In 1952, L\'evy published a very influential paper\cite{L52}, dealing 
with the description of nuclear forces by means of pion exchanges.
The basic $\p N$ interaction was assumed to be pseudoscalar,
{\em "since it is only in this case that the intrinsic field theoretical 
infinities  can be separated and re-interpreted consistently"},
and several Feynman diagrams describing the two-body system 
were calculated.
In a subsequent work, which appeared in 1953, Klein\cite{K53} corrected some 
of L\'evy's results and considered the role of three-body forces.
He also adopted pseudoscalar $\p N$ coupling and a pattern that would
become important later began to appear, namely that dominant contributions
were associated with diagrams containing $N \bar{N}$ intermediate states.
Another remark that would have future impact was hidden in a 
note added in proof:
{\em "Numerical calculations show that the potential obtained in the 
paper does not agree with experiment.
Further work, to be published, indicates moreover, that the 
perturbation theory doesn't even converge."} 

Almost simultaneously, Drell and Huang\cite{DH53} were investigating the
problem of nuclear saturation, which requires a mechanism for short range 
repulsion.
In their abstract, they state:
{\em "[...]The leading term in the $n$-body potential depends only on the 
interparticle distance and is repulsive (attractive) for $n$ odd (even)."}
This finding is materialized in their interesting fig.6 and 
the special role of three-body forces is stressed in the
concluding section:
{\em "[...]we find that many-body forces, and in particular the three-body
repulsion, provide a satisfactory qualitative understanding of nuclear
saturation."}

The qualitative findings by both Klein and Drell-Huang
were confirmed and summarized by Wentzel\cite{W53}:
{\em "In pseudoscalar meson theory ($PS$ coupling), the saturation character 
of nuclear forces is accounted for by the cooperative action of many-body
forces. In particular, the repulsive 3-, 5-, 7-, ... body forces keep the
nucleus from collapsing into a very small volume.
These nuclear interactions result mainly from matrix elements describing
virtual nucleon-pair creation and annihilation processes[...]"}. 

These conclusions were disputed, already in 1953, by Brueckner and 
Watson\cite{BW53}.
In their own words:
{\em 
"We have seen that it is possible to derive a nucleon-nucleon potential,
working entirely with a nonrelativistic approximation to the pseudoscalar meson 
theory.
[...]
We have depended rather strongly on the suppression of nucleon pair 
formation by radiative effects; the contributions to the potential 
from pair formation in high order calculated without taking into
account such effects otherwise tend to be so large as to invalidate 
the power series expansions usually used."}
This perception underlies their ad hoc calculation procedure,
which became known as {\bf pair suppression} and was widely adopted. 
In another publication along this line, by 
Brueckner, Levinson and Mahmoud\cite{BLM54}, one finds:
{\em "The application of these techniques to a specific problem of 
considerable interest, namely the two-body potentials of pseudoscalar
meson theory (adjusted to fit the low-energy scattering parameters)
shows that these potentials also have the necessary characteristics
to give an approximately correct description of nuclear saturation."}
In the same paper these authors also argue that many-body forces
produce negligible effects.

This intense debate about nuclear forces raised many problems regarding 
the applicability of ordinary perturbation theory to strong interactions. 
As a consequence, in the mid-fifties, lagrangians had fallen into disgrace
and theoretical interest shifted to the analytic structure of the S-matrix,
investigated by means of dispersion relation techniques.  
The rehabilitation of strong lagrangians had to wait for the development 
of chiral symmetry.

%--- 5 ----------------------------------------------------------------------
\section{The first modern three-body force}

\begin{figure}[h]
\includegraphics[width=.7\columnwidth,angle=0]{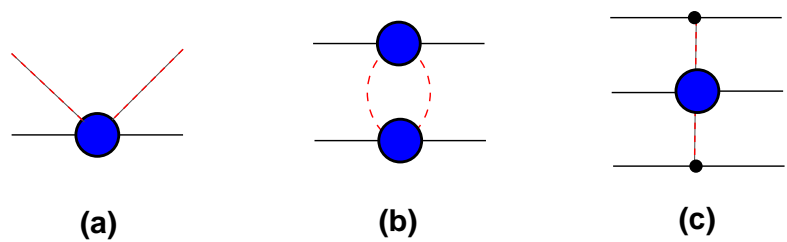}
\caption{(Color online) (a) Free $\p N$ amplitude;
(b) two-pion exchange two-body potential; 
(c) two-pion exchange three-body potential.} 
\label{F1}
\end{figure}

In the second half of the fifties, the idea of deriving potentials from 
first principles was put aside and replaced by the more pragmatic 
attitude of blending theory and empirical information.
At that time, the relationship between the off-shell $\p N$ amplitude
and important components of the potential, as indicated in fig. \ref{F1},
had already been suggested.
Also, many measurements of elastic $\p N$ scattering, 
performed in the same period, had led to a considerable improvement 
of the empirical information about the on-shell amplitude.
If suitably treated by means of dispersion relations, a rather valuable
information about the off-shell $\p N$ amplitude could be obtained
from empirical data.
\vspace{2mm}

% --- fig.1 ^^^^^^^^^^^^^^^^^^^^^^^^^^^^^^^^^^^^^^^^^^^^^^^^^^^^^^^^^^^^^^^^^^
%^^^^^^^^^^^^^^^^^^^^^^^^^^^^^^^^^^^^^^^^^^^^^^^^^^^^^^^^^^^^^^^^^^^^^^^^^^^^^

At the end of 1956, a paper incorporating the new attitude was published 
by Miyazawa\cite{M56}.
Its title was 
{\em "Interaction of $P$- and $S$-Wave Pions with Fixed Nucleons"}
and, in the abstract, one finds the clear statement:
{\em "A method is given of replacing pion scattering parts in a Feynman 
diagram by experimentally observed quantities.
[...]
Two examples are given: (1) The anomalous magnetic moment of the proton
is rigorously expressed in terms of pion-nucleon scattering amplitudes [...].
(2) The internucleon potential is also expressed by means of scattering
quantities. 
In this case the number of virtual pions exchanged between the two nucleons
is limited to two, although the number of pions emitted and absorbed by the
same nucleon is not limited."} 

In the cases of $P$- and $S$-waves, the scattering matrix elements 
for the process $\p^i(k) N \rar \p^j(q ) N$ were written respectively as 
\bea
\la j,q|S|i, k \ra \!\!\!\!&=&\!\!\!\! 2\p i \d(k_0 \sm q_0)\,
\lb A(k_0)\; \tau_i \tau_j \,\bsig\cd\bk \; \bsig\cd\bq
+ B(k_0) \, \lp \tau_i \tau_j \,\bsig\cd\bq \; \bsig\cd\bk
+ \tau_j \tau_i \,\bsig\cd\bk \; \bsig\cd\bq \rp \right.
\nn\\[1mm]
&+&\!\!\!\! \left. C(k_0) \tau_j \tau_i \,\bsig\cd\bq \; \bsig\cd\bk \rb
e^{i(\bk\sm\bq)\cdot\bx} \;,
\label{e.4}\\[2mm]
\la j,q|S_s|i, k \ra \!\!\!\!&=&\!\!\!\! 2\p i \d(k_0 \sm q_0)\,
\lb D(k_0)\; \tau_i \tau_j 
+ E(k_0) \; \tau_j \tau_i\rb
e^{i(\bk\sm\bq)\cdot\bx} \;,
\label{e.5}
\eea

\ni
and dispersion relations were used to derive the functions $A$, $B$ and $C$
from empirical total cross sections, whereas $D$ and $E$ were obtained
from scattering lengths. 
The importance of these results lies in the fact that they hold for both 
real and virtual pions.
The derivation of the two-pion exchange component 
of the $NN$ potential, fig. 1b, was performed with great care, 
so as to avoid various possibilities of double counting.

In 1957, Fujita and Miyazawa published the work 
{\em "Pion Theory of Three-Body Forces"}\cite{FM57}, 
in which the off-shell $\p N$ amplitude also plays an essential role, 
as in fig. 1c.
Their construction of the two-pion exchange three-nucleon potential 
$(TPE$-$3NP)$ is described:
{\em  "The process being considered is this. 
Particle (1) emits a pion which is scattered by particle (2) and then
absorbed by particle (3).
[...]
For the scattering by particle (2), experimental values can be used.
Actually, this is a scattering of a virtual pion having zero energy.
However, the use of dispersion relations makes it possible to correlate
this scattering with real scatterings.
In this way all virtual transitions on particle (2) are correctly 
taken into account."} 

This clear understanding of the role played by the off-shell $\p N$ 
amplitude in the $TPE$-$3NP$ establishes the work of Fujita and Miyazawa 
as the first one in the modern tradition.
With variations, the essence of their approach is present in the leading
terms of potentials constructed ever since. 

On the more technical side, an explicit analytic form for the potential 
was presented in terms of Yukawa functions and numerical values for the 
constants $A(0)$, $B(0)$ and $D(0)$ were derived from $\p N$ scattering data.
A little of manipulation allows their original expressions to be recast 
in the form of $TPE$-$3NP$ adopted nowadays, given by
\bea
&& V_L(123) = -\,\frac{\m}{(4\p)^2} \, 
\lc \d_{ab} \lb a \; \m - b \; \m^3 \, \bnb_{12} \cdot \bnb_{23} \rb 
+ d\; \m^3 \; i\,\e_{bac} \tau_c^{(2)}\;
i \, \bsig^{(2)} \cdot \bnb_{12} \times \bnb_{23}\rc
\nn\\[2mm]
&& \;\;\;\;\; \times 
\lp (g_A \, \m/2 \,f_\p)\;\tau_a^{(1)} \;\bsig^{(1)} \cdot \bnb_{12} \rp\;
\lp (g_A \, \m/2 \,f_\p)\;\tau_b^{(3)} \;\bsig^{(3)} \cdot \bnb_{23} \rp\;
Y(x_{12}) \; Y(x_{23}) \;,
\label{e.6}
\eea

\ni
where $\m$ is the pion mass and $a$, $b$ and $d$ are strength constants.
The first one is associated with $S$-waves and the other two, 
with $P$-waves.
The interpretation of this result is rather simple.
The term within curly brackets describes the off-shell $\p N$ scattering 
on nucleon 2, those within parentheses represent the emission 
and absorption of a single pion on nucleons 1 and 3,
whereas the Yukawa functions $Y$ describe pion propagation.
To my knowledge, this structure for the potential was first obtained
by Fujita and Miyazawa, indicating the power and generality of their approach.

A final comment about their work is in order. 
It is well known that $\D$ intermediate states play a dominant role in 
some low-energy $\p N$ cross sections and, by extension,
the same happens in the three-body force.
On the other hand, the structure of the actual Fujita-Miyazawa potential
also incorporates other effects and is rather general. 
Therefore the widespread identification of the F-M force with $\D$
intermediate states corresponds, at least, to an undue over-simplification
of their results.

The off-shell $\p N$ amplitude derived in ref.\cite{M56} was also used
in the study of more involved components of the three-nucleon force,
by Fujita, Kawai and Tanifuji\cite{FKT62}, in 1962.
In that work, particular attention was paid to both the so called ring 
diagrams and to processes containing exchanges of one pion with one of 
the nucleons and two pions with the other, 
which continue to be object of attention nowadays.

%--- 6 ----------------------------------------------------------------------
\section{the age of chiral symmetry}

The discovery, in 1956, of parity non-conservation in weak interactions,
motivated a great interest about the nature of weak currents.
The well known electromagnetic current  is represented by a vector $(J^\m)$ 
and conserves parity.
In the case of weak interactions, on the other hand, one finds both
vector $(V^\m)$ and axial-vector $(A^\m)$ currents, which respectively 
conserve and do not conserve parity.
The electromagnetic current is conserved $(\dr \cd J=0)$, but 
the same does not happen with the weak ones. 
Nevertheless, in the late fifties, the notion was developed that, 
in some special limits, weak currents could be considered as 
being approximately conserved.
This was translated into approximate symmetries of the interaction lagrangian,
associated with transformations that can be schematically represented as 
\beq
[V,V]\rar V, \;\;\;\;\;\;
[V,A]\rar A, \;\;\;\;\;\;
[A,A]\rar V\;.
\label{e.7}
\eeq 

Loosely speaking, changes in parity occur in images produced by mirrors 
and one may say that the action of an axial operator is analogous
to transforming a right hand into a left one. 
A vector operator does not change parity and corresponds to transforming
a right hand into itself. 
This kind of analogy led the transformations indicated in eqs.(\ref{e.7})
to be called {\bf chiral}, a word derived from {\bf hand} in Greek.

In 1960, {\bf chiral symmetry}, implemented by means of the linear-sigma 
model\cite{GL60}, has been applied with great success
to strong interactions.
This model assumes the existence of a scalar-isoscalar 
particle called $\s$ and proved to have a very rich structure. 
One of its beautiful features is the picture of a strong vacuum,
which is both covariant and not empty.

\begin{figure}[hbt]
%\vspace{-5mm}
%\hspace*{-25mm}
\includegraphics[width=.9\columnwidth,angle=0]{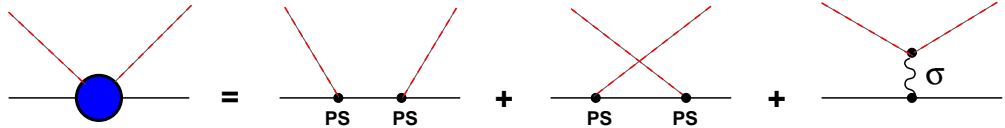}
%\vspace{-50mm}
\caption{(Color online) Structure of the $\p N$ amplitude 
in the linear-sigma model .} 
\label{F2}
\end{figure}

Another important result was the natural explanation produced 
for the observed smallness of the $\p N$ scattering lengths,
which puzzled research in the early fifties.
This achievement is crucial for Nuclear Physics.
The outer layers of nucleon interactions are due to pion exchanges
and many important components of the force depend on the intermediate
$\p N$ amplitudes shown in fig. 1. 
In the $\s$-model, the $\p N$ amplitude is given by the
three diagrams shown in fig. \ref{F2}, with pseudoscalar $\p N$ coupling.
The first two diagrams correspond to the model adopted in the early 
fifties\cite{L52} and explicit calculation of their contributions yields
$(a_{PS}^+, a_{PS}^-)=(-1.84, 0.136) \m^{-1}$, where the superscripts 
$\pm$ refer to isospin channels.
The comparison of these results with the experimental values\cite{H83} 
$(a_{exp}^+, a_{exp}^-)=(-0.008, 0.092) \m^{-1}$ shows that the 
prediction for $a^+$ is too large by a factor of about 200. 
The third diagram, containing the exchange of the $\s$, is the signature
of chiral symmetry in this process and produces 
$(a_\s^+, a_\s^-)=(1.83, 0) \m^{-1}$.
A large cancellation occurs in 
$(a_{PS}^+ \sp a_\s^+, a_{PS}^- \sp a_\s^-)=(-0.01, 0.136) \m^{-1}$, 
turning both sums compatible with empirical orders of magnitude.
Related cancellations also happen in the case of nuclear forces and a 
pedagogical discussion can be found in ref.\cite{BRR97}.

\vspace{2mm}
% --- fig.2 ^^^^^^^^^^^^^^^^^^^^^^^^^^^^^^^^^^^^^^^^^^^^^^^^^^^^^^^^^^^^^^^^^^
%^^^^^^^^^^^^^^^^^^^^^^^^^^^^^^^^^^^^^^^^^^^^^^^^^^^^^^^^^^^^^^^^^^^^^^^^^^^^^

\begin{figure}[h]
\includegraphics[width=0.9\columnwidth,angle=0]{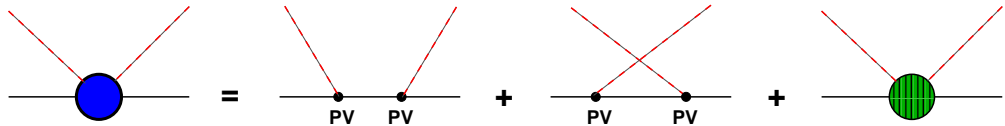}
\caption{(Color online) Chiral tructure of the $\p N$ amplitude; 
the dashed (green) bubble represents a polynomial contribution.} 
\label{F3}
\end{figure}

Two remarks about the chiral $\p N$ amplitude are important at this point.
The first one is that the smallness of the full $\s$-model result
is a direct consequence of chiral symmetry.
The linear $\s$-model corresponds to just one among many possibilities for 
implementing the symmetry in hadronic systems and the order of magnitude 
of the chiral $\p N$ amplitude is independent of the method employed.
The second remark is that the symmetry, powerful as it is, cannot predict
the full empirical content of the $\p N$ interaction.
At low-energies, empirical information about this process is usually 
encoded into a polynomial in the variables $\nu$ and $t$, proposed by 
H\"ohler and collaborators\cite{H83}.
The coefficients of this polynomial are obtained by extrapolating experimental
information to the region below threshold, by means of dispersion relations.
For this reason, they are called {\bf subthreshold coefficients}.
So, in order to construct a $\p N$ amplitude suitable to be employed in 
nuclear interactions, one uses chiral symmetry, supplemented by 
subthreshold information, as indicated in fig. \ref{F3}. 

\vspace{2mm}
% --- fig.3 ^^^^^^^^^^^^^^^^^^^^^^^^^^^^^^^^^^^^^^^^^^^^^^^^^^^^^^^^^^^^^^^^^^
%^^^^^^^^^^^^^^^^^^^^^^^^^^^^^^^^^^^^^^^^^^^^^^^^^^^^^^^^^^^^^^^^^^^^^^^^^^^^^

The incorporation of chiral symmetry into hadronic amplitudes is 
performed rigorously and many results have the status of {\bf theorems}.
In the simplest versions, these theorems are exact for low-energy interactions
of massless pions.
The masses of actual pions, although small, are non-vanishing and it is
in this sense that chiral symmetry is approximate.
However, the transposition of chiral theorems to the case of 
massive pions is also performed rigorously and, as a consequence, 
predictions from the approximate symmetry are both under control 
and unambiguous.

In general, chiral theorems or amplitudes have the form of power 
series\footnote{This series also includes chiral logarithms an other 
non-analytic terms.} 
in a typical scale $q$, set by either pion four-momenta or nucleon 
three-momenta, such that $q\ll 1$ GeV.
As power series\footnote{In a Taylor series such as 
$\sin x \simeq x - x^3/3! + \cdots$, the r.h.s. is obtained by means 
of a well defined set of operations over the l.h.s.$=\sin x$.
On the other hand, in the case of a chiral amplitude $A_\chi$, 
one has schematically
$A_\chi \simeq A_L \, q^L + A_{NL}\, q^{L\sp 1} + \cdots$
and the coefficients $A_i$ must be determined by means of field theory,
since one does not know the l.h.s.},
these theorems involve both leading order terms and corrections.
The former have been evaluated for a variety of problems since the sixties, 
whereas the latter is th object of the {\bf chiral perturbation theory},
which gained momentum in the last two decades.

%--- 7 ----------------------------------------------------------------------
\section{the leading order three-body force}

The leading chiral three-body force begins at $\cO(q^3)$ and is due to
the process shown in fig.1c.
It took a long time between the first derivations
were performed and a kind of consensus about the uniqueness of the
result could be reached by practitioners of the field.

The first application of chiral symmetry to three-body forces
appeared in 1968, in a paper by Brown, Green and Gerace\cite{BGG68},
in the case of nuclear matter.
They state in the abstract:
{\em "[...]
it is shown that long-range three-body forces in nuclear matter are 
essentially zero.
This means that they are also small in nuclei."}
This conclusion was soon disputed in a letter by McKellar and 
Rajaraman\cite{MKR68}, where one finds:
{\em "[...]
we present below two models [...], both of which indicate that three-body
forces in nuclei are about as strong as the two-body forces."} 
A compromise between these extremes was reached in a third paper, by
Brown and Green\cite{BG69}, whose abstract begins with:
{\em "An earlier claim about the smallness of three-body effects in nuclear 
matter is examined in the light of a criticism.
Our conclusion is that, whereas the lowest-order three-body effect is 
small, second order effects are larger,[...]".} 

The next step was given in 1974, by Yang\cite{Y74},who shifted attention
to the triton and was the first to use the so called {\bf chiral dynamics}.
As he explains in the abstract:
{\em "We have derived a two-pion exchange three-body potential using a chiral
invariant Lagrangian for the pion-nucleon interaction.
Only the three dominant lowest order processes which give rise to a 
three-body interaction are included: 
(a) The second pion is emitted before the first pion is absorbed and the 
nucleon which emits the second pion goes into a positive-energy state; 
(b) one nucleon exchanges a $\rho$ meson with a pion;
(c) one nucleon is scattered into an $N^*(1236)$."}
In the main text, one finds an instructive description of how 
effective lagrangians were striking back.
After emphasizing the role of the off-shell intermediate $\p N$ amplitude, 
he says:
{\em 
"The PCAC (partially conserved axial-vector current) and the current algebra
developed in the late 1960's have been quite successful in describing
the soft-pion process, e,g., the Adler's consistency conditions$^4$ on 
strong interactions and low-energy parameters$^5$ of the $\p N$ scattering
are found to be in good agreement with the experiments.
It was noted$^6$ that for soft-pion processes the predictions obtained 
with current algebra can also be derived by a different method:
Just use the lowest-order graphs generated by a chiral-invariant Lagrangian.
We shall apply this method[...]".}
His reference $^6$ is a famous paper by Weinberg\cite{W67}.
Another passage shows how chiral symmetry 
was displacing the old pair-suppression mechanism of the fifties:
{\em 
"The three-body force arising from a virtual nucleon-antinucleon pair
[...] is suppressed, since the soft-pion theory tells us explicitly
that $T^{\mathrm{Born}}$ is computed according to the gradient coupling
scheme."}
Approximations done in the treatment of process (a), however, does not 
allow his results to be cast in the form of eq.(\ref{e.6}).

Just a little later, in 1975, another paper focused on nuclear matter
appeared, by Coon, Scadron and Barrett\cite{CSB75}.
Their strategy for dealing with dynamics is presented in the abstract:
{\em 
"We re-examine the off-shell $\p N$ amplitude occurring in the two-pion
exchange three-body force, subject to all the constraints of current
algebra.
This amplitude is not dominated by the $\D(1231)$ isobar;
instead, if the $\s-$term is known, it can be determined from on-shell
scattering." }
In the third section, {\em "Off-shell pion-nucleon scattering"},
one finds a detailed description of their approach, as well
as a comparison with results form other works. 
With hindsight, this work can be considered as being a precursor of 
the Tucson-Melbourne (TM) potential\cite{TM79}, published in 1979.
In this later paper, previous results were generalized and
its abstract reads:
{\em "We derive the complete three-nucleon potential of the two-pion exchange
type, suitable for nuclear structure calculations, by extending away from
the forward direction the subthreshold off-pion-mass-shell $\p N$ amplitude
of Coon, Scadron and Barrett.
The off-mass-shell extrapolation, subject to current algebra and PCAC
constraints, yields approximately model independent amplitudes
(in that they depend primarily on $\p N$ data) in the complete potential.
[...]".}
This version of the $3NP$ was restricted to a particular reference
frame, but this condition was lifted a little later\cite{CG81}.
In those papers, the notation used in eq.(\ref{e.6}) was 
introduced.
The TM force also included another structure, represented by a parameter
$c$ within the curly bracket, which was written as 
$\{ \d_{ab} [ a \; \m - b \; \m^3 \, \bnb_{12} \cdot \bnb_{23} 
- c\; \m^3 \, (\bnb_{12}^2 +\bnb_{23}^2)] 
+ d\; \m^3 \; i\,\e_{bac} \tau_c^{(2)}\;
i \, \bsig^{(2)} \cdot \bnb_{12} \times \bnb_{23} \}$.
This term proved to be problematic in numerical calculations and 
has been identified with short range effects\cite{Rsr,tame}.
It was eventually 
removed\footnote{For furhter details, the reader is 
directed to Prof. Coon's contribution to this volume.}
in a paper by Coon and Han\cite{CH01}, published in 2001,
where parameters for a new version, known as TM', were
presented.  

In 1983, the $TPE$-$3NP$ known as Brazil potential was 
produced\footnote{The last term of eq.(67) in that paper 
contains a misprinted sign, which was corrected in ref.\cite{tame}.},
by Coelho, Das and Robilotta\cite{CDR83}.
It was based on effective lagrangians and relied on a chiral model 
for the $\p N$ amplitude derived by Olsson and 
Osypowski\cite{OO} which included, besides the same $\rho$ and $\D$ 
intermediate processes considered by Yang\cite{Y74},
a parametrized form for the $S$-wave interaction. 
In that model, all the parameters were tuned to empirical subthreshold 
coefficients and the resulting structure conveyed model
independent information.
This gave rise to a potential in agreement with eq.(\ref{e.6}). 
In 1986, the possibility that the long range physics 
in both the TM and Brazil forces could be contaminated
by the inclusion of form factors, required by numerical calculations,
was discussed\cite{tame}.

It is worth recalling that chiral expansions are performed by means of 
well defined theoretical rules and hence their results must be unique.
This holds, in particular, for leading terms.
In the case of three-body forces, the leading term has the 
generic structure given by eq.(\ref{e.6}), whereas free parameters
are determined by subhtreshold coefficients. 
A sample of values adopted for potential parameters along five decades
is given in table \ref{T1}.
At present, variations in those values reflect just the
empirical information used as input and are totally unrelated with chiral 
symmetry. 
\vspace{2mm}
\begin{table}[h]
\begin{tabular}{ccccccc}
\hline
\tablehead{1}{r}{b}{ref.}
&\tablehead{1}{r}{b}{year}
& \tablehead{1}{r}{b}{potential}
& \tablehead{1}{c}{c}{$a\m$}
& \tablehead{1}{c}{c}{$b\m^3$}
& \tablehead{1}{c}{c}{$c\m^3$} 
& \tablehead{1}{c}{c}{$d\m^3$}  \\
\hline
\cite{FM57} & 1957 & F-M & -0.27 & -1.24 & 0     & -0.31 \\
\cite{TM79} & 1979 & TM  & 1.13   & -2.58  & -1.05 & -0.75  \\
\cite{CDR83} & 1983 & BR  & -1.05  & -2.29  & 0     & -0.77  \\
\cite{CH01} & 2001 & TM'-93 & -0.74 & -2.53& 0     & -0.72  \\
\cite{CH01} & 2001 & TM'-99 & -1.12 & -2.80& 0     & -0.75  \\
\hline
\end{tabular}
\caption{Leading-order parameters for the $TPE$-$3NP$ (dimensionless units).}
\label{T1}
\end{table}

%--- 8 ----------------------------------------------------------------------
\section{Chiral perturbation theory}

Chiral perturbation theory (ChPT) is the art of deriving systematic
{\bf corrections} to leading order terms in chiral expansions. 
In the case of Nuclear Physics, the research program was outlined 
by Weinberg\cite{W9091} and one of his papers deals specifically with
three-body forces\cite{W92}.
In ChPT, emphasis is put on a lagrangian which possesses
approximate chiral symmetry and is written as a string of terms with 
different orders in the power-counting parameter $q$.
The coefficients of higher order terms are known as  {\bf low-energy constants} 
(LECs) and their presence is felt in final results for chiral amplitudes.
The values of the LECs cannot be fixed by the symmetry and
must be determined from experimental information.
In the case of nuclear potentials, the basic source of information are
the usual $\p N$ subthreshold coefficients.
For this reason, the result for the {\bf leading} term, which was already 
unique and given by eq.(\ref{e.6}) before ChPT, continues to be so afterwards, 
in spite possible apparent changes in form due to rephrasing.
On the other hand, as stated in the first paper to deal with applications
of  ChPT, by Ord\'o\~nez and van Kolck\cite{OvK92}, in 1992, 
the new method has the clear advantage of stressing the model 
independence of the results. 
ChPT is also important for preventing whimsical diagramatics in
complex problems. 

From an operational point of view, the power of ChPT in nuclear forces 
is that it allows the generation of new structures within a unified 
theoretical framework. 
One possible direction for new developments concerns {\bf short-range}
physics.
In the case of three-body forces, short-range chiral interactions
were considered in 1994 by van Kolck\cite{vK94} and by 
Friar, Huber, and van Kolck\cite{FHvK99}, in 1999.
In the later 
paper\footnote{One notes that the values quoted for the F-M and 
Brazil forces in table II of ref.\cite{FHvK99} are in qualitative 
disagreement with those given in table \ref{T1} of the present work.},
the abstract points to the solution of an old problem:
{\em "It is demonstrated that the short-range $c$ term of the Tucson-Melbourne
force is unnatural in terms of power counting an should be dropped."}

The second possible direction is related with the production of 
new mathematical structures, which go beyond eq.(\ref{e.6}).
At present, next-to-leading order corrections to the three-body force, 
which involve a rather large number of diagrams, are being calculated
by Epelbaum, Meissner and collaborators\cite{EM07}.
The reader is directed to the contribution by Prof. Meissner to 
this volume for a comprehensive account of their results.  

A taste of what is about to come can be felt in a recent work by
Ishikawa and Robilotta\cite{IR07}, in which a selected subset of 
corrections to the leading term, associated with the long-range part 
of the $TPE$-$3NP$, was considered.
The final result has the structure  
$V(123)=V_L(123)+ [V_{\d L}(123) + \d V(123)]$, 
where $V_L$ is the old leading term given by eq.(\ref{e.6}),
whereas the factor within square brackets includes corrections 
due to ChPT.
The term $V_{\d L}$ can be obtained directly from $V_L$, by replacing 
$(a,b,c)$ with $(\d a,\d b, \d c)$.
In other words, this part of the correction corresponds to 
shifts in the parameters of the leading term,
which are numerically smaller than $10\%$.
The factor $\d V(123)$, on the other hand, represents effects described 
by new mathematical functions, whose actual forms are too cumbersome
to be displayed here. 
For the present purposes, it suffices to say that they involve both non-local 
operators and the replacement of the Yukawa functions in eq.(\ref{e.6})
by more complicated propagators involving loop integrals. 
The strengths  of these new functions are described by a set of new
parameters $e_i$, which are also typically smaller than $10\%$ of the 
leading ones.
So, the impact of ChPT in table \ref{T1} is to produce both small
modification in already existing coefficients and the 
appearance of many new columns.
The latter are by far the most interesting ones, since they may contain the
explanation for effects such as the $A_y$ puzzle\footnote{Discussed by
Prof. Tornow in this volume.}.

%--- 8 ----------------------------------------------------------------------
\section{concluding remarks}

Progress in our understanding of the world always involves both 
conceptual ruptures and continuities.
In this work, I have tried to recall that this feature was also present
in the development of three-body forces, a highly collective task.
If one compares the discussions of strong interactions held in the
fifties with those occurring nowadays, one may be tempted to stress
discontinuity.
On the other hand, reflecting about the work by Fujita and Miyazawa,
one notices that their formulation of the problem in terms of dispersion
relations corresponds more to a bending of its path than to a 
complete break with the past.
It is therefore best described as a {\bf turning point}.
Chiral symmetry came next and has promoted a solid understanding
of both meanings and values of the strength parameters of the force.
And it is remarkable that chiral perturbation theory, 
which guides the present wave of research, is not replacing, 
but rather extending and complementing the results
derived fifty years ago by Fujita and Miyazawa.

%--- A ----------------------------------------------------------------------   
\begin{theacknowledgments}
It was a great pleasure participating in the FM50 Symposium and I
would like to thank the organizers for this nice opportunity.
\end{theacknowledgments}

%^^^^^^^^^^^^^^^^^^^^^^^^^^^^^^^^^^^^^^^^^^^^^^^^^^^^^^^^^^^^^^^^^^^^^^^^^^^
%BBBBBBBBBBBBBBBBBBBBBBBBBBBBBBBBBBBBBBBBBBBBBBBBBBBBBBBBBBBBBBBBBBBBBBBBBBB

\end{document}